\documentclass[10pt]{iopart}
\usepackage{epsfig}
\def\Journal#1#2#3#4{{#1} {\bf #2}, #3 (#4)}
\def\ZPC{{\em Z. Phys.} C}

\begin{document}
\noindent

\title[Survey of experimental data]{Survey of experimental data}
\author{H. Oeschler
}
\address{
Institut f\"ur Kernphysik, Darmstadt University of Technology, \\
D-64289 Darmstadt, Germany\\
}
\begin{abstract}
A review of meson emission in heavy ion collisions at incident energies
from SIS up to collider energies is presented.
A statistical model assuming chemical equilibrium and local strangeness 
conservation (i.e.~strangeness conservation per collision)
explains most of the observed features.

Emphasis is put onto the study of $K^+$ and $K^-$ emission at low incident
energies. 
In the framework of this statistical model it is shown that the experimentally
observed equality of $K^+$ and $K^-$ rates at ``threshold-corrected'' energies
$\sqrt{s} - \sqrt{s_{th}}$ is due to a crossing of two excitation functions.
Furthermore, the independence of the $K^+/K^-$ ratio on the number of
participating nucleons observed between SIS  and RHIC is consistent with
this model.

It is demonstrated that the $K^-$ production at SIS energies occurs
predominantly via strangeness exchange and that this channel
is approaching chemical equilibrium.
The observed maximum in the $K^+/\pi^+$ excitation function 
is also seen in the ratio of strange to non-strange particle production.
The appearance of this maximum around 30 $A\cdot$GeV
is due to the energy dependence of the chemical freeze-out parameters
$T$ and $\mu_B$.  
\end{abstract}

\section{Introduction}
Central heavy ion collisions
at relativistic incident energies represent an ideal tool to study
nuclear matter at high temperatures.
Particle production is -- at all incident energies -- a key quantity to extract
information on the properties of nuclear matter under these extreme conditions.
Particles carrying strangeness have turned out to be very valuable messengers.

A specific purpose of this paper is the presentation of the evolution
of strange particle production over a large range of incident energies.
Many results are shown together with a theoretical interpretation.
The attempts to describe particle production yields with 
statistical models~[1-8]
have turned out to be very successful over this large domain of incident 
energies.

\section{General Trends}

\subsection{Production of pions and kaons from SIS to RHIC}

\begin{figure}
\begin{minipage}[t]{5.3cm}
\epsfig{width=7.1cm,file=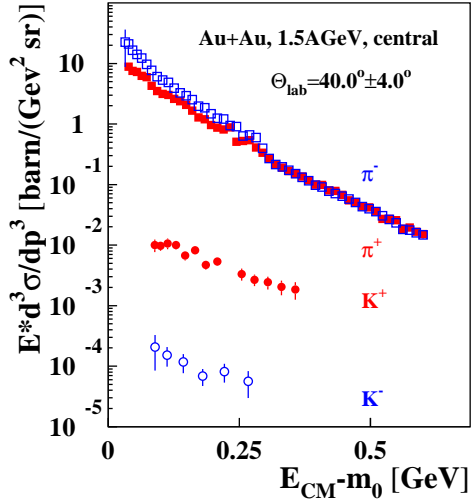}
\end{minipage}\hfill
\begin{minipage}[b]{7.8cm}
\vspace*{-3cm}
\caption{Spectra of positively and negatively
charged pions and kaons measured in central collisions of Au+Au 
at 1.5 $A\cdot$GeV. {\it Preliminary results.}}
\label{spectra_au}\end{minipage}\end{figure}

At incident energies around 1 $A\cdot$GeV pion and kaon production is very
different: Pions can be produced by direct $NN$ collisions in contrast to kaons.
The threshold for $K^+$ production in $NN$ collisions is 1.58 GeV
and only collective effects can accumulate the energy needed to produce
a $K^+$ together with a $\Lambda$ 
due to strangeness conservation.
The threshold for $K^-$ production is even higher (2.5 GeV)
as they are produced as $K^+$ $K^-$ pairs. These conditions lead to very 
different yields for the various mesons as demonstrated in 
Fig.~\ref{spectra_au} showing some spectra from
central Au+Au collisions at 1.5 $A\cdot$GeV.
The yields of pions are much higher than for kaons; positively and negatively
charged pions differ due to the $N/Z$ ratio of Au. The yield of $K^+$ is 
by nearly two orders of magnitude higher than the one of $K^-$ due to the 
different thresholds. At higher incident energies all these differences diminish
as will be discussed.

\begin{figure}\begin{minipage}[t]{4.8cm}
\epsfig{width=7.6cm,file=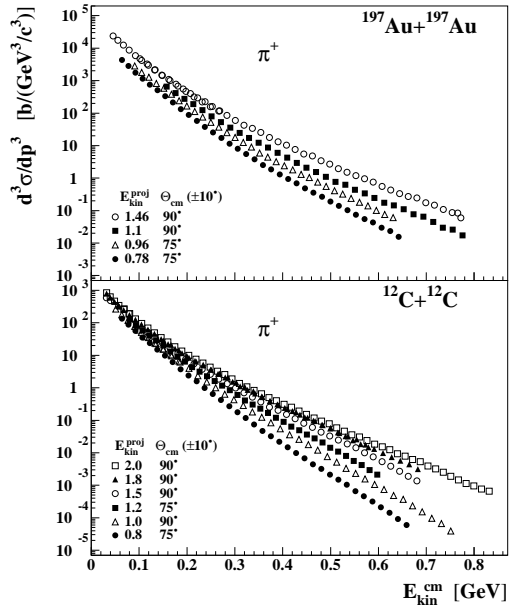}
\end{minipage}\hfill\begin{minipage}[b]{7.8cm} 
\caption{Spectra of positively charged pions in the center-of-mass frame
close to midrapidity for Au+Au (upper part) and
C+C (lower part).
A deviation from a Boltzmann shape is seen at all incident energies and for both
collision systems.}
\label{pion_p}\end{minipage}\end{figure}

Pion spectra of excellent quality are now available or will appear 
soon~[9-10]. 
Figure~\ref{pion_p} shows as an example spectra 
of positively charged pions as $d^3\sigma/dp^3$ (Boltzmann representation)
for C+C and Au+Au collisions at different incident energies.
The spectra are measured close to mid rapidity.

All spectra exhibitconcave shapes in this representation
deviating from a Boltzmann distribution which would be a straight line. 
Yet, even in a thermal condition one is not expecting a pure Boltzmann 
distribution as pions originate both from ``free'' pions and from 
resonance decay after freeze out. 
This type of shape is observed up to the highest incident energies and the 
above interpretation holds qualitatively. For a quantitative description with 
these two components still some work has to be done.


                                    
\begin{figure}\hspace*{-.1cm}
\epsfig{width=7.3cm,file=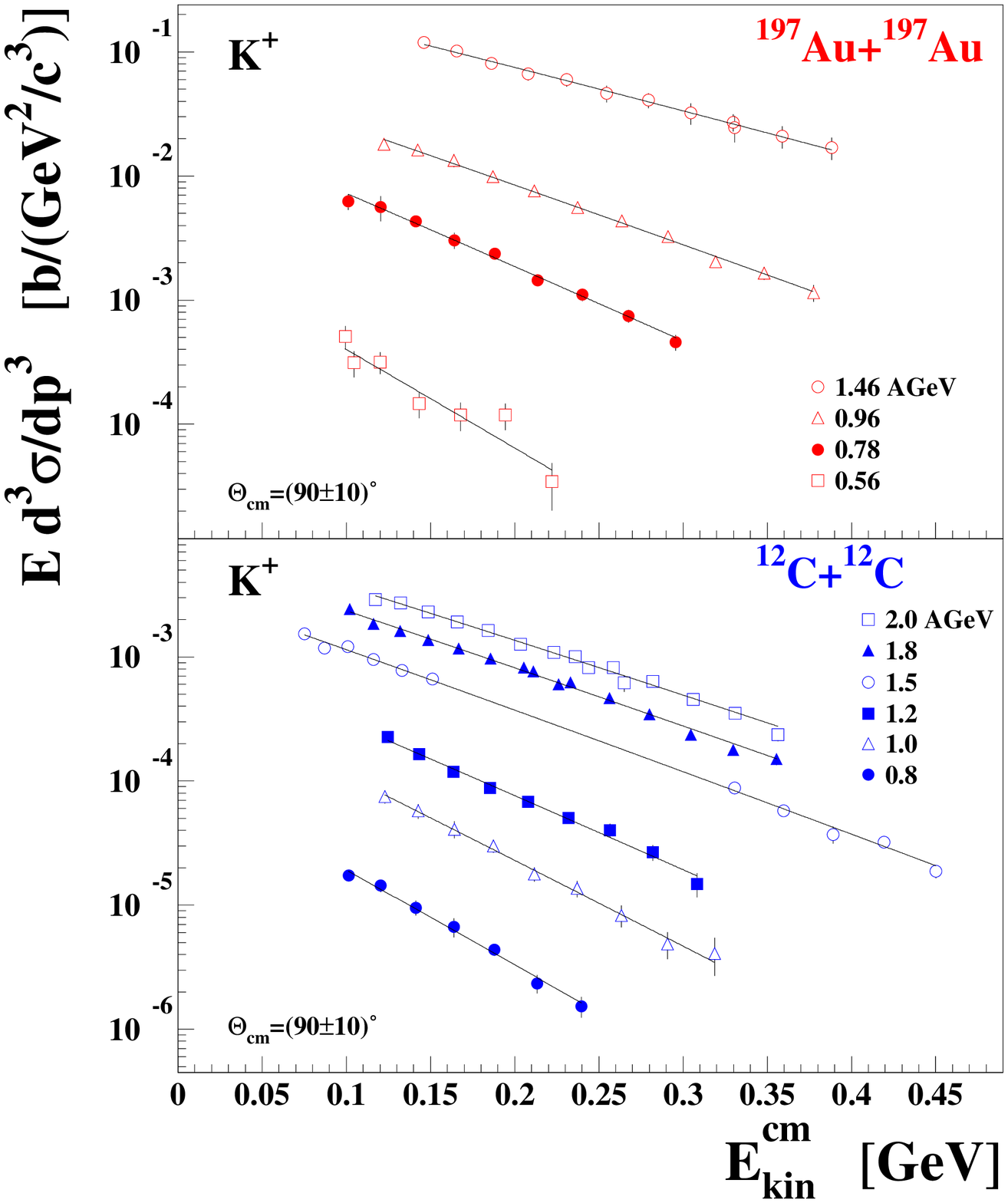}
\hspace*{-.6cm}
\epsfig{width=7.3cm,file=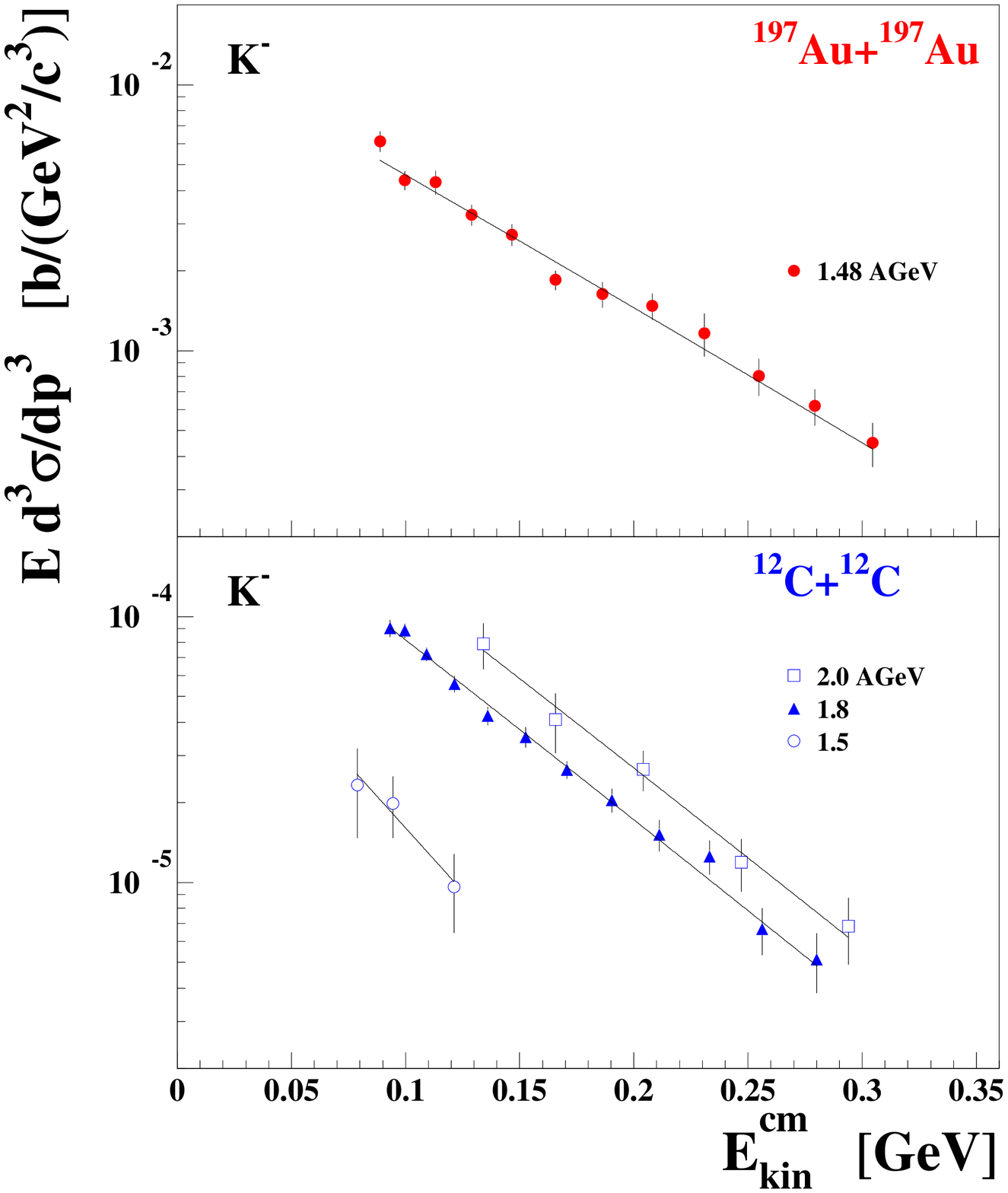}
\caption{$K^+$ and $K^-$ spectra measured at midrapidity for Au+Au (upper part)
and C+C (lower part) at various incident energies~\cite{Laue, CS}.
The spectra of $K^-$ from Au+Au collisions are preliminary.}
\label{KP_KM}\end{figure}

Spectra of $K^+$ and $K^-$ from mass-symmetric systems C+C and Au+Au
at incident energies from 0.6 to 2.0 $A\cdot$GeV
measured at midrapidity are shown in Fig.~\ref{KP_KM}.
They exhibit Boltzmann shapes in all cases~\cite{Laue, CS}.
Their inverse slope parameters increase
monotoneously with incident energy
and the heavier system exhibits harder spectra than the light system at the
same incident energy.
                                   
\begin{figure} 
\begin{minipage}[t]{5.3cm}
\mbox{\epsfig{width=7.9cm,file=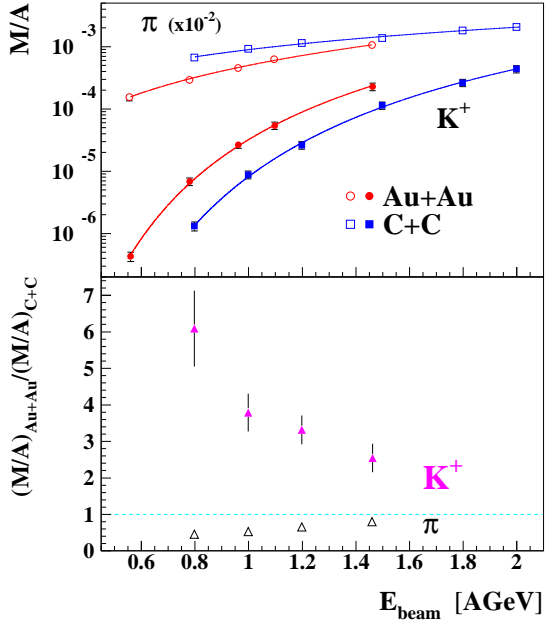}}
\vspace*{-3cm}
\end{minipage}\hfill
\begin{minipage}[b]{6.9cm}
\caption{Upper part: Multiplicities of $K^+$ and pions per $A$
as a function of incident energy.
The pion data include charged and neutral pions.
The lines represent fits to the data (see~\cite{CS}).
Lower part: Ratio of the multiplicities per $A$ (Au+Au over C+C) as a function
of incident energy.}
\label{EOS}\end{minipage}\end{figure}

The interaction of $K^+$ and $K^-$ with nuclear matter is very different:
Due to their $\bar s$ content $K^+$ cannot be absorbed, while $K^-$
can easily be absorbed on a nucleon converting it into a $\Lambda$.
This difference makes the $K^+$ to be messengers of the early stage of 
the collision.
Therefore $K^+$ are ideal probes for this dense stage and allow to extract
the stiffness of the nuclear equation of state~\cite{CS}. The basis of these
studies is the ratio of $K^+$ measured in Au+Au and in C+C collisions
as shown in Fig.~\ref{EOS}. This subject is presented in detail in the talk by
C.~Sturm. 

\begin{figure}[h]\begin{minipage}[t]{5.3cm}
\mbox{\epsfig{width=9cm,file=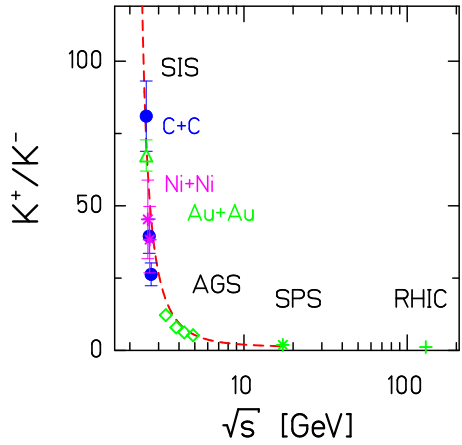}}
\end{minipage}\hfill
\begin{minipage}[b]{7.6cm}
\caption{Ratio of positively and negatively
charged kaons for various collisions systems from SIS up to RHIC.
{\it Preliminary results.}
The dashed line represents the results of calculations using the statistical
model.}
\label{kp_kmm_ratio}
\end{minipage}\end{figure}

Figure~\ref{KP_KM} demonstrates that the yield of $K^+$ is much higher than
the one of $K^-$ as mentioned already. This is caused by the different
production thresholds for $K^+$ and $K^-$. 
At AGS energies the ratio of $K^+/K^-$ has decreased
to about 5 \cite{Ahle} and is as low as 1.16 at RHIC~\cite{Star}. 
This trend is summarized in Fig.~\ref{kp_kmm_ratio}.
The dashed line represents the results of calculations using a statistical 
model~\cite{CLE99, CLE00}.

\subsection{Interpretation within a statistical model}

\begin{figure} [h]
\begin{minipage}[t]{5.3cm}
\mbox{\epsfig{width=9.0cm,file=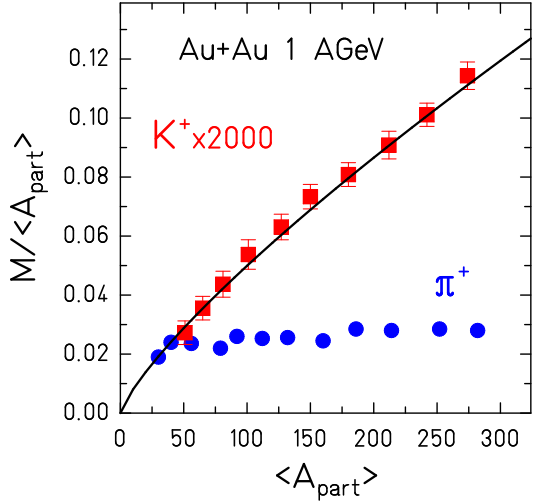}}
\end{minipage}
\hfill
\begin{minipage}[b]{7.6cm}
\vspace*{-3.6cm}
\caption{The multiplicity of $K^+/A_{part}$ rises strongly
with $A_{part}$ in contrast to the pion multiplicity~\protect\cite{Mang}.
This rise can be described by the statistical model
including local strangeness conservation (see text).}
\label{Apart}
\end{minipage}
\end{figure}

Pions and $K^+$ exhibit a further 
very pronounced contrast: While the pion multiplicity per number of
participating nucleons $A_{part}$ remains constant 
with $A_{part}$, the $K^+$ 
multiplicity per $A_{part}$ rises strongly (Fig.~\ref{Apart}).
The latter observation seems to be in conflict with a thermal
interpretation, which -- in a naive view -- should give multiplicities per
mass number $A$ being constant.

Usually, the particle number densities or
the multiplicities per $A_{part}$, here for pions, are described
in a simplified way by a Boltzmann factor
$$\frac{M_{\pi}}{A_{part}}\sim \exp \left(-\frac{<E_{\pi}>}{T}\right),
$$with the temperature $T$ and the total energy $<E_{\pi}>$.

The production of strange particles has to fulfil strangeness
conservation. The attempt to describe the measured particle ratios
including strange hadrons at AGS and SPS using a strangeness chemical potential
$\mu_S$ is quite successful~\cite{Cley,PBM,pbm99,PBM_RHIC}.
However, this grand-canonical treatement is not sufficient
if the number of produced strange particles is small.
Then a statistical model has to take care of {\it local
strangeness conservation} in each reaction as introduced in
~\cite{Hagedorn}.
This is done by taking into account
that e.g.~together with each $K^+$ a $\Lambda$ or another strange particle
is produced:
$$\frac{M_{K^+}}{A_{part}}\sim \exp \left(-\frac{<E_{K^+}>}{T}\right)
\left[g_{\Lambda}V \int {d^3p\over (2\pi)^3}
\exp\left(-{{(E_{\Lambda}-\mu_B)}\over T}\right)\right],
$$where $T$ is the temperature, $\mu_B$ the baryo-chemical potential,
$g_i$ the degeneracy factors, $V$ the production volume for making the
associate pair (see~\cite{CLE99,CLE00}) and $E_i$ the total energies.
We note that this volume is not identical to the volume of the system at freeze
out.
The volume parameter $V$ is
taken as $r^3_V A_{part}$ with a common $r_V$ = 1.07 fm for all systems and all
incident energies.

This formula, simplified for demonstration purposes,
neglects other combinations leading to the production
of $K^+$ as well as the use of Bose-Fermi distributions, which are all
included in the computation.
The corresponding formula for $K^-$ production 
$$\frac{M_{K^-}}{A_{part}}\sim \exp \left(-\frac{<E_{K^-}>}{T}\right)
\left[g_{K^+}V \int {d^3p\over (2\pi)^3}
\exp\left(-{E_{K^+}\over T}\right)\right].
$$is similar, but does not
depend on $\mu_B$. This point will become important later on.

These formulae lead to
a reduction of $K^+$ and $K^-$ yields as compared to the numbers calculated
without exact strangeness conservation~\cite{CLE99,CLE00}.
Two extreme conditions can be seen from these equations.
In the limit of a small number of strange particles
the additional term (due to the parameter $V$)
leads to a linear rise of $M_{K^+}/A_{part}$,
while $M_{\pi}/A_{part}$ remains constant.
This is in very good agreement with the experimental
observations shown in Fig.~\ref{Apart}. 
For very high temperatures or very large volumina, the terms in brackets
approach unity (see Ref.~\cite{CLE99}) resulting in
the grand-canonical formulation. This is much better seen in the exact
formulae using modified Bessel functions~\cite{CLE99,CLE00,Hamieh}.

At low incident energies, the particle ratios (except $\eta/\pi_0$)
are well described using this canonical approach~\cite{CLE99}.
Surprisingly, even the measured $K^+/K^-$ ratio is described and 
this ratio does not depend on the choice of the volume term $V$.
It should be noted that the statistical model uses nominal masses of the
particles while some transport calculations \cite{Cassing}
have to reduce the $K^-$ mass (as expected for kaon in the nuclear medium)
in order to describe the measured yields. 

\begin{figure}[h]\begin{minipage}[t]{5.1cm}
\mbox{\epsfig{width=8.0cm,file=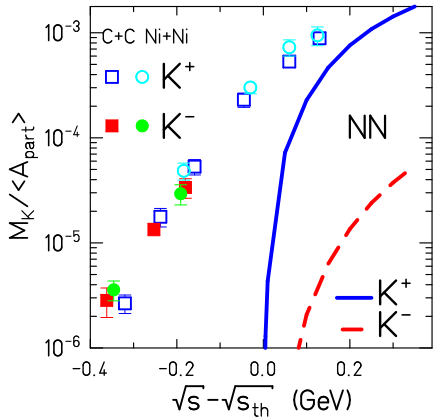}}
\vspace*{-3cm}\end{minipage}\hfill
\begin{minipage}[b]{6.9cm}\vspace*{-3cm}
\caption{Measured $K^+$ and $K^-$ yields in heavy ion and $NN$ collisions
as a function of  $\sqrt{s}-\sqrt{s_{th}}$~\protect{\cite{Laue,Barth,Marc}}.
$<A_{part}>$ is $A/2$ for heavy ion data and 2 for $NN$ collisions.}
\label{KP_KM_sthr}\end{minipage}\end{figure}

Before comparing the data with the calculations in detail, 
a summary of the measurements by the KaoS Collaboration is given.
These results have attracted considerable interest as in heavy ion collisions
the $K^-$ yield compared to the $K^+$ cross section
is much higher than expected from $NN$ collisions \cite{Laue, Barth}.
This is especially evident if the kaon multiplicities are
plotted as a function of $\sqrt{s} - \sqrt{s_{th}}$ where
$\sqrt{s_{th}}$ is the energy needed to produce the respective
particle in $NN$ collisions
taking into account the mass of the associately produced partner.
To produce a $K^+$ in $NN$ collisions $\sqrt{s_{th}}$ = 2.548 GeV
and a $K^-$ $\sqrt{s_{th}}$ = 2.87 GeV.
The obvious contrast between $NN$ and $AA$ collisions, shown in 
Fig.~\ref{KP_KM_sthr}, has lead to the interpretation of the results by
in-medium properties which cause e.g.~a lower threshold for
$K^-$ production when produced in dense matter~\cite{Cassing}.
The observed difference between $NN$ and $AA$ collisions alone
is not sufficient to conclude on properties of
kaons in matter. In heavy ion collisions, kaons can be produced by other
channels, e.g.~$\pi \Lambda \rightarrow K^- N$ which are not available in $NN$
collisions. Only by using detailed transport-model calculations
one might conclude on new properties of kaons in matter~\cite{Cassing}.

\begin{figure}[h]
\begin{minipage}[t]{5.1cm}
\mbox{\epsfig{width=9.0cm,file=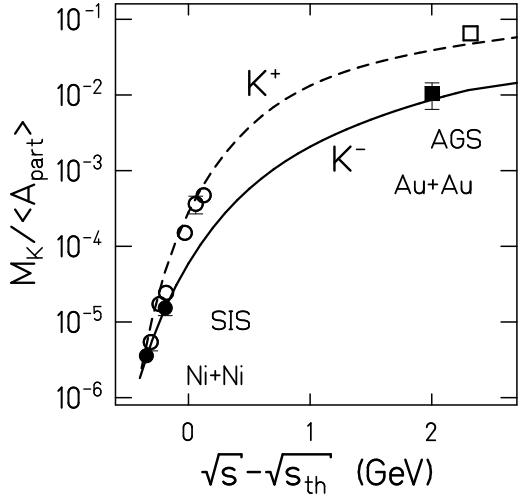}}
\end{minipage}
\hfill
\begin{minipage}[b]{7.8cm}
\vspace*{-3.8cm}
\caption{Calculated $K^+/A_{part}$ and $K^-/A_{part}$ ratios
in the statistical model as a function of $\sqrt{s} - \sqrt{s_{th}}$
for Ni+Ni collisions.
The points are results for Ni+Ni collisions at SIS energies
\protect\cite{Barth,Marc}
and Au+Au at 10.2 $A\cdot$GeV (AGS)~\protect\cite{Ahle}. 
At AGS energies the influence of the system mass is negligible.}
\label{KP_KM_sthr_therm}
\end{minipage}
\end{figure}

It is therefore of interest to see how the results of the statistical model
appear in a representation where the $K^+$ and $K^-$ multiplicities are given
as a function of $\sqrt{s} - \sqrt{s_{th}}$.
Figure~\ref{KP_KM_sthr_therm} demonstrates that at
values of $\sqrt{s} - \sqrt{s_{th}}$ less than zero
the excitation functions for $K^+$ and $K^-$ cross leading to 
the observed equality of 
$K^+$ and $K^-$ at SIS energies.
The yields differ at AGS energies by a factor of five.
The difference in the rise of the two excitation functions
can be understood by the formulae given above.
The one for $K^+$ production contains ($E_{\Lambda}-\mu_B$) while the other
has $E_{K^+}$ in the exponent of the second term. 
As these two values are different, the
excitation functions, i.e.~the variation with $T$, exhibit a different rise.

\begin{figure}[h]
\begin{minipage}[t]{5.3cm}
\mbox{\epsfig{width=9.1cm,file=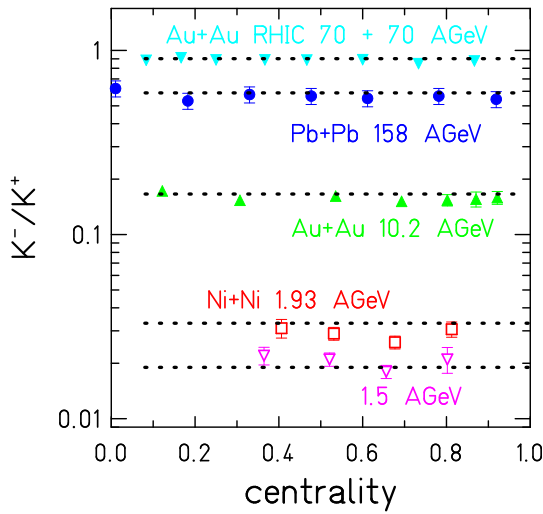}}
\end{minipage}
\hfill
\begin{minipage}[b]{7.5cm}
\vspace*{-5.8cm}
\caption{The $K^+/K^-$ ratio appears to be constant of a function of centrality
from SIS up to RHIC energies. The dotted lines represent the predictions of 
the statistical model. Data from \protect\cite{Ahle,Star,Marc}.}
\label{KPKM_SIS_RHIC}
\end{minipage}
\end{figure}

Furthermore, the two formulae predict that the $K^+/K^-$ ratio for a given 
collision should not vary with centrality as $V$ cancels in the ratio.
This has indeed been observed in Au+Au/Pb+Pb collisions between 1.5 $A\cdot$GeV
and RHIC energies~\cite{AF,Star,Marc,Ahle,Dunlop}
as shown in Fig.~\ref{KPKM_SIS_RHIC}.
This independence of centrality is most astonishing
as one expects at low incident energies an influence
of the different thresholds and the density variation with centrality.
For instance at 1.93 $A\cdot$GeV the $K^+$ production is above and
the $K^-$ production below their respective $NN$ thresholds.
             
Transport-model calculations show clearly that
strangeness equilibration requires a time interval of
40 -- 80 fm/$c$~\cite{Koch86,Brat00}.
On the other hand statistical models assuming chemical equilibration
are quite successful in describing the particle yields
including strange particles.

In the case of $K^+$ production,
no strong absorptive channel seems to be available which could lead to 
chemical equlibration.  
For $K^-$ production the situation is quite different.
At low incident energies strange quarks are found only in
a few hadrons. The $\bar s$ quark is essentially only in $K^+$, while
the $s$ quark will be shared between $K^-$ and $\Lambda$ (or other hyperons).
This sharing of the $s$ quark might be in chemical equlibrium as
the reactions $$\pi^0 + \Lambda \rightleftharpoons p + K^- \quad \rm{or} \quad
\pi^- + \Lambda \rightleftharpoons n + K^-$$
 are strong and have only
slightly negative Q-values of -176 MeV.

The idea that the $K^-$ yield is dominated by strangeness exchange via
the $\pi^- + \Lambda$ channel has been suggested by~\cite{Ko84}
and has been demonstrated quantitavely in a recent theoretical
study~\cite{Hart01}. The direct $K^+ K^-$ pair production via baryon-baryon
collisions has negligible influence as these $K^-$ are absorbed entirely.
In these transport-model calculations the strangeness exchange is approaching
equilibrium but does not fully reach it~\cite{Hart01}. For details see the 
talk given by C.~Hartnack. 

If these reactions are the dominating channels, they might reach chemical 
equilibration, e.g.~the rates for producing $K^-$ and for absorbing $K^-$
are equal. Then the law of mass action
can be applied giving for the respective concentrations~\cite{oeschler_s2000}
$$\frac{[\pi^{0,-}] \cdot [\Lambda]}{[K^-] \cdot N} \, = \, \kappa .$$
As the number of $K^-$ relative to $\Lambda$ is small, $[\Lambda]$ can be
approximated by $[K^+]$ and rewriting gives
$$\frac{[K^-]}{[K^+]} \propto M(\pi^0 + \pi^-)/A_{part}.$$

This relation also explains that the measured ratio of $K^-/K^+$ is constant 
with centrality (Fig.~\ref{KPKM_SIS_RHIC}) as the pion multiplicity does not 
vary with centrality.
\begin{figure}[h] 
\mbox{\epsfig{width=4.2cm,file=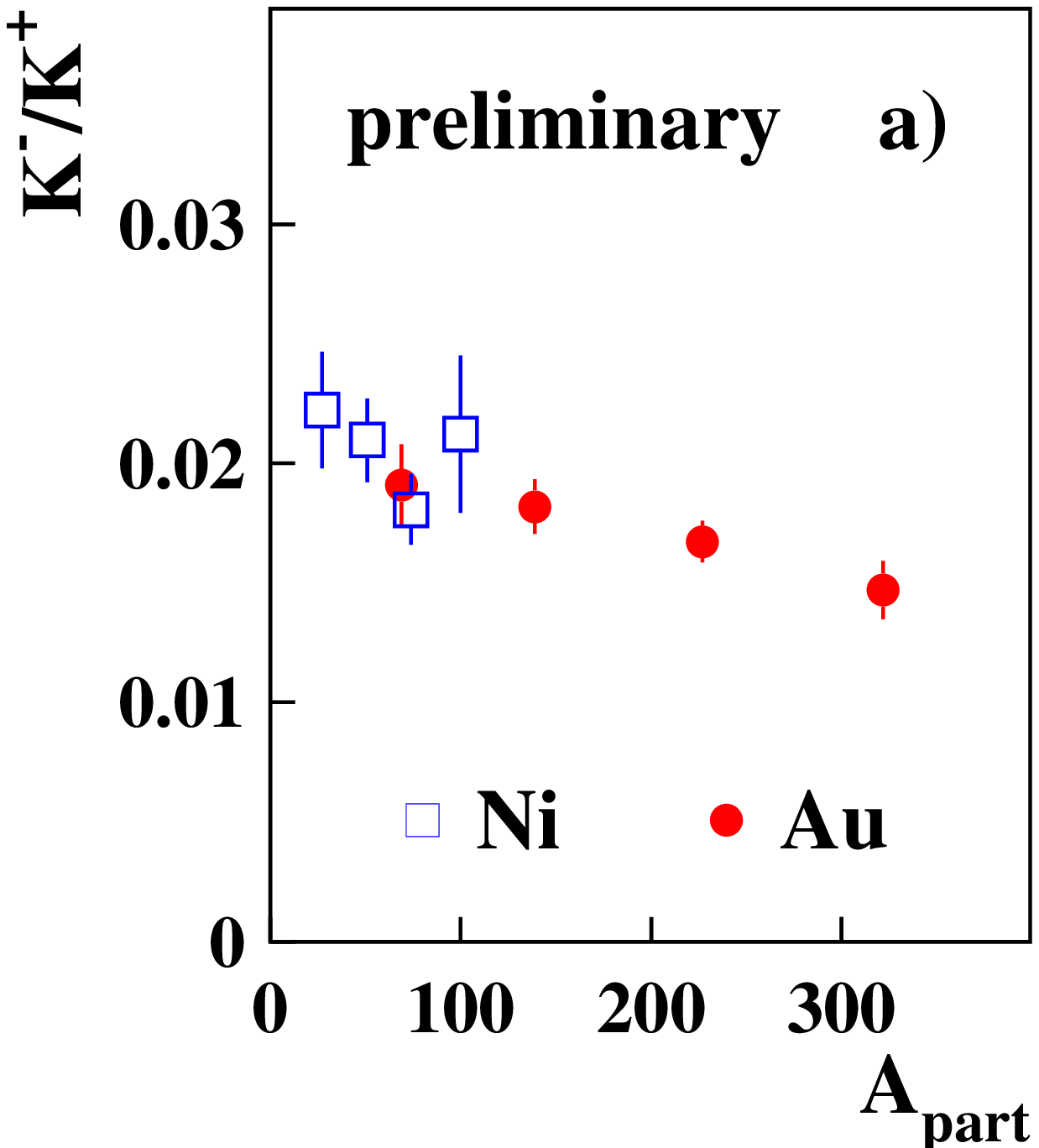}}
\mbox{\epsfig{width=4.2cm,file=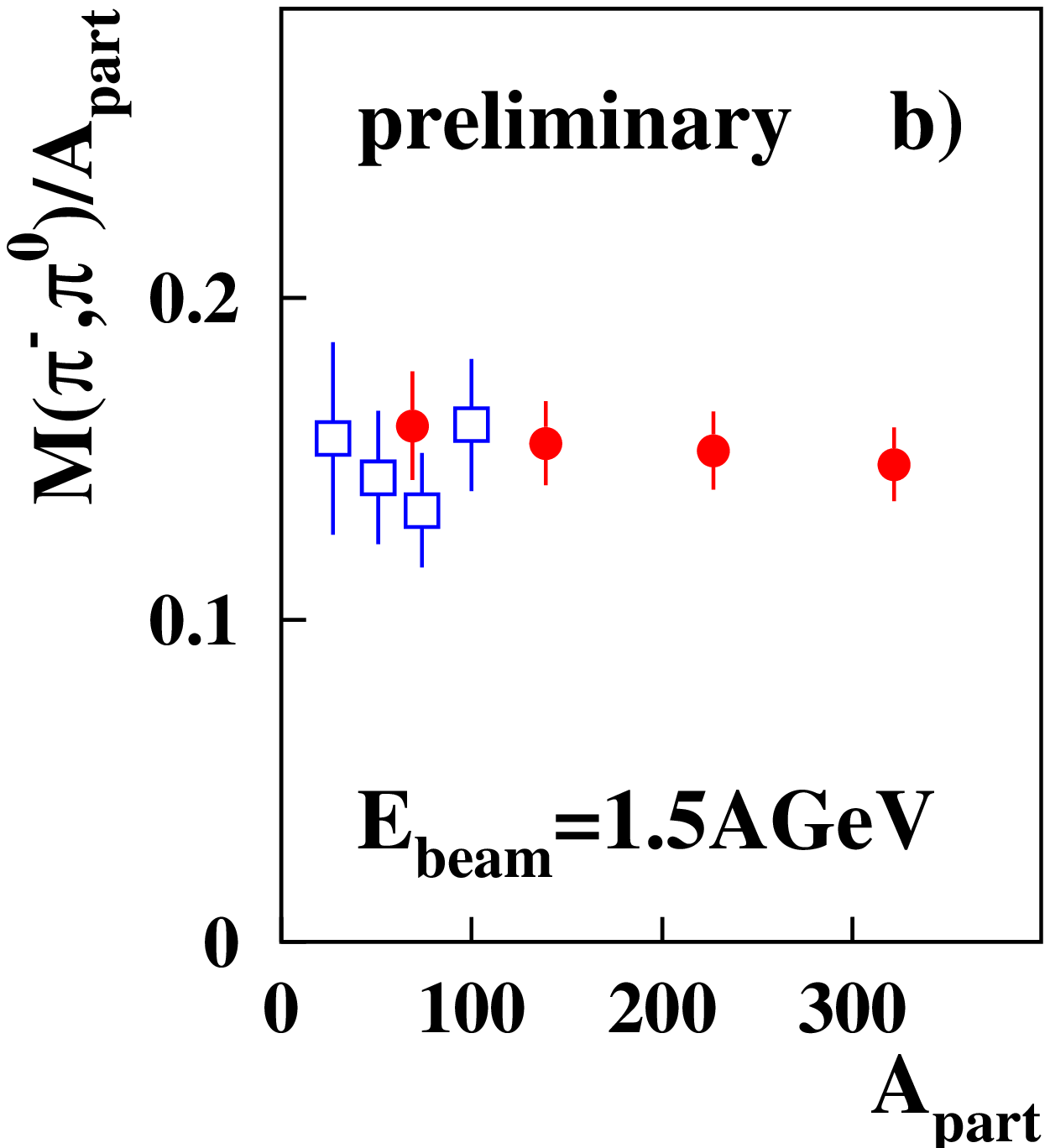}}
\mbox{\epsfig{width=4.2cm,file=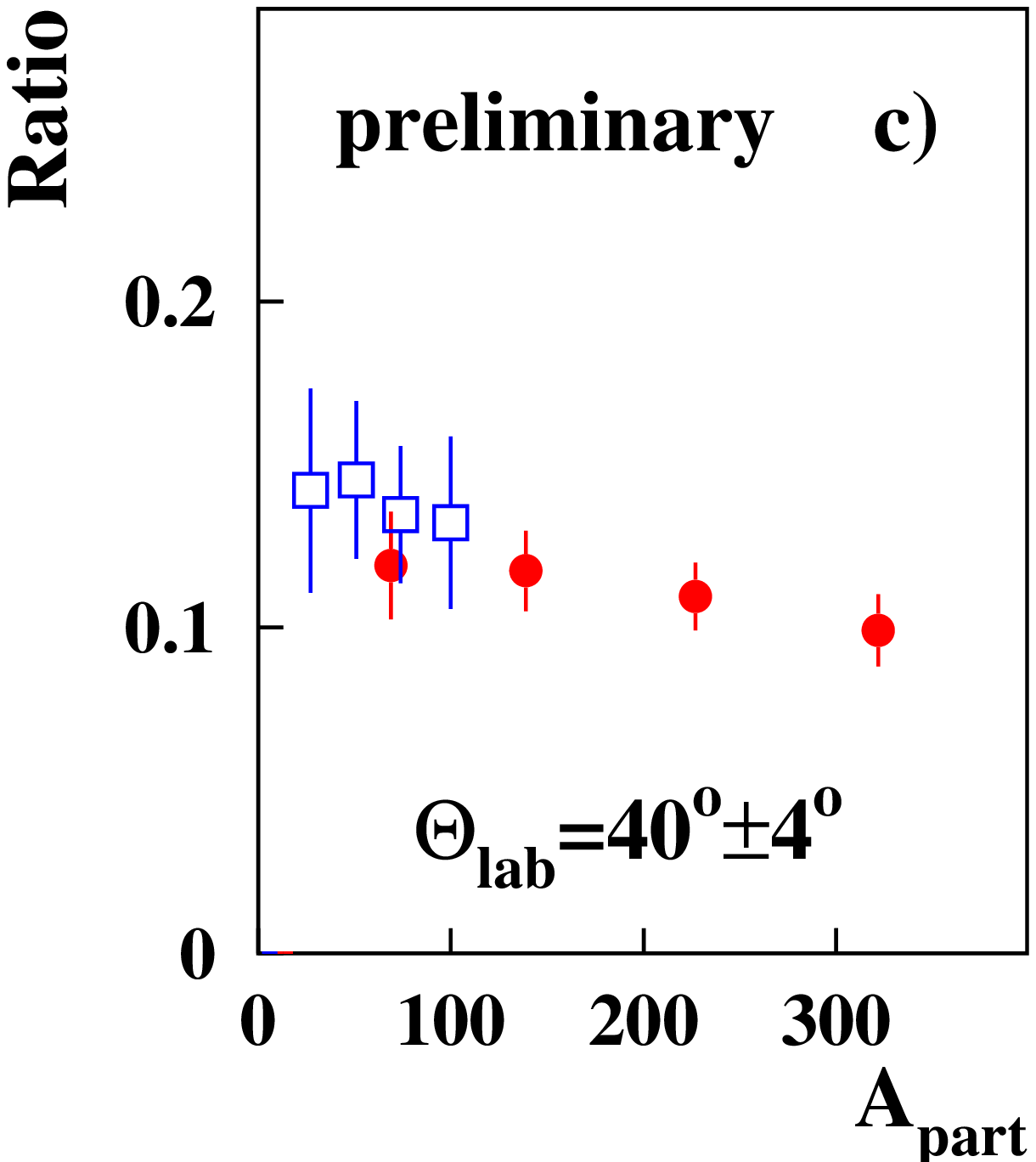}}
\caption{Measured~$K^-/K^+$ ratio, $M(\pi^0 + \pi^-)/A_{part}$ and~the double
ratio $([K^-]/[K^+])/([M(\pi^0 + \pi^-)]/A_{part})$  as a function of
$A_{part})$ both for Ni+Ni and Au+Au collisions at 1.5 $A\cdot$GeV.
{\it Preliminary results.}}
\label{Massaction}
\end{figure}
A detailed study for the low energies has been given in the talk 
by A.~F\"orster.
Figure~\ref{Massaction} summarises by demonstrating the constancy 
of the $K^-/K^+$ ratio and of the pion multiplicity with $A_{part}$ 
for Ni+Ni and Au+Au collisions at 1.5 $A\cdot$GeV~\cite{AF,Marc,FU}.
It turns out that these ratios do not even
depend on the choice of the collision system. 
The right part of this figure exhibits the double ratio
$([K^-]/[K^+])/([M(\pi^0 + \pi^-)]/A_{part})$ which shows only a minor 
deviation from a horizontal line. This result can be taken as an argument that
this specific channel is not far from chemical equilibrium.

\begin{figure}[h]
\begin{minipage}[t]{5.3cm}
\mbox{\epsfig{width=9.2cm,file=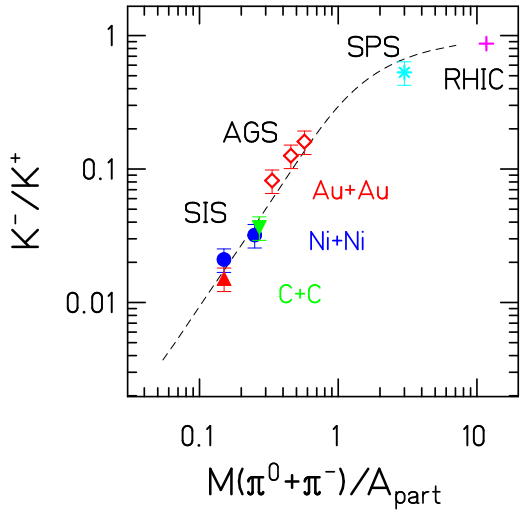}}
\end{minipage}
\hfill
\begin{minipage}[b]{7.8cm}
\vspace*{-5cm}
\caption{The $K^-/K^+$ ratio as a function of the
pion multiplicity $M(\pi^-+\pi^0)/A_{part}$ as a test of the law of mass action.
{\it Preliminary data.} The dashed line shows the prediction of the statistical
model.}
\label{Massaction_SIS_RHIC}
\end{minipage}
\end{figure}

Next we test in Fig.~\ref{Massaction_SIS_RHIC} 
the validity of the law of mass action by plotting the
$K^-/K^+$ ratio as a function of the pion multiplicity 
$M(\pi^0 + \pi^-)/A_{part}$ at incident energies from SIS up to RHIC.
At SIS and AGS energies the direct relation holds, i.e.~the $K^-/K^+$ ratio
rises with $M(\pi^0 + \pi^-)/A_{part}$. 
At SPS and RHIC energies the linear relation is no longer valid;
$K^-$ are obviously produced by other channels, 
i.e.~$K^+ K^-$ pair production. This change of the dominating channel is well
reproduced by the statistical model (dashed line in 
Fig.~\ref{Massaction_SIS_RHIC}).

\subsection{Maximum relative strangeness content
 in heavy ion collisions around 30 A$\cdot$GeV} 

\begin{figure}[h]
\begin{minipage}[t]{5.3cm}
\mbox{\epsfig{width=7.0cm,file=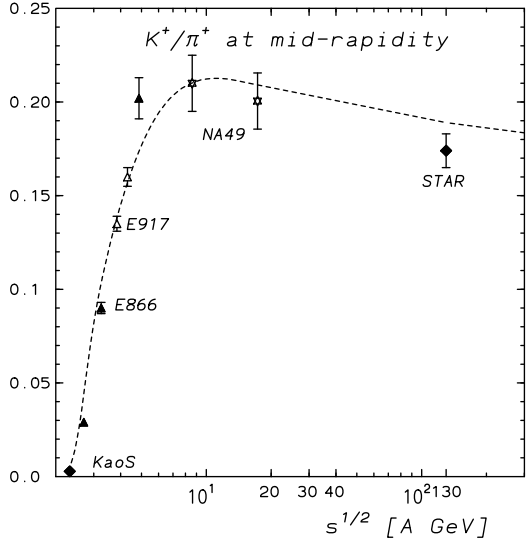}}
\end{minipage}
\hfill
\begin{minipage}[b]{7.8cm}
\vspace*{-3.6cm}
\caption{$K^+/\pi^+$ ratio obtained around midrapidity as a
function of $\sqrt s$ from the various experiments. 
The dashed
line shows the results of the statistical model in complete equilibrium.}
\label{KP_PI_ratio}
\end{minipage}
\end{figure}

The experimental data from heavy ion collisions show that
the $K^+/\pi^+$ ratio rises from SIS up to AGS but it is larger
for AGS than at the highest CERN-SPS energies
\cite{CLEY98,Ahle,Dunlop,blume,bearden} 
and decreases even further at RHIC \cite{Star} as shown in 
Fig.~\ref{KP_PI_ratio}.

 This behavior is of
particular interest as it could signal the appearance of new
dynamics for strangeness production in high energy collisions. It
was  even conjectured~\cite{gazdzicki}
 that this property could
indicate   an energy    threshold  for  quark-gluon plasma
formation in relativistic heavy ion collisions.
Some transport models are able to describe the occurrence of this 
maximum within a continuous evolution of hadron rescattering 
and string degrees of freedom~\cite{Wang}.

In the following we
analyze the energy dependence of strange to non-strange
particle  ratios in the framework of a hadronic
statistical model. In the whole  energy range, the
hadronic yields  observed in heavy ion collisions resemble those
of a population in chemical equilibrium along a unified freeze-out
curve determined by the condition of fixed energy/particle
$\simeq$ 1 GeV \cite{CLEY98} providing a relation between 
the temperature $T$ and the baryon chemical potential $\mu_B$.
As the beam energy increases $T$ rises and $\mu_B$ is slightly reduced.
Above AGS energies $T$ exhibits only a
moderate change and converges to
its maximal value in the range of 160 to 180 MeV, while $\mu_B$ is strongly 
decreasing.

Instead of studying the $K^+/\pi^+$ ratio we use the ratios of
 strange to non-strange particle
multiplicities (Wroblewski
factor)~\cite{wroblewski}
defined as
$$
\lambda_s \equiv {2\bigl<s\bar{s}\bigr>\over
\bigl<u\bar{u}\bigr> + \bigl<d\bar{d}\bigr>}
$$
where the  quantities in angular brackets refer to the number of
newly formed quark-antiquark pairs, i.e.~it excludes all
quarks that were present in the target and the projectile.

\begin{figure}
\begin{minipage}[t]{5.0cm}
\mbox{\epsfig{width=8.5cm,file=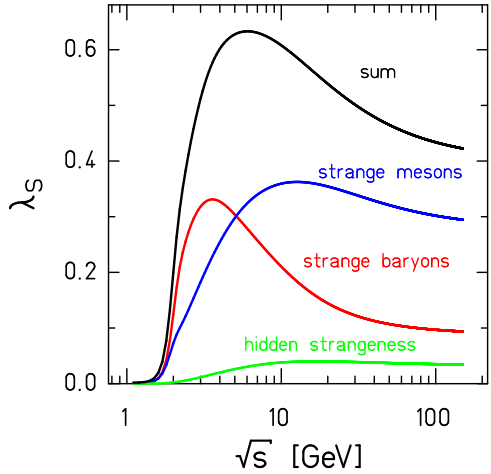}}
\end{minipage}
\hfill
\begin{minipage}[b]{7.8cm}
\vspace*{-4.9cm}
\caption{Contributions to the Wroblewski factor $\lambda_s$ (for definition
see text) from strange baryons, 
strange mesons, and mesons with hidden strangeness. 
The sum of all contributions is given by the full line.}
\label{Wrob_composition}
\end{minipage}
\end{figure}

Applying the statistical model to particle production in heavy ion
collisions calls for the  use of the canonical ensemble
to treat the number of strange particles
particularly  for data in the energy range
from SIS up to AGS \cite{CLE99,ko} as mentioned before.
The calculations for Au-Au and
Pb-Pb collisions  are performed using
a canonical correlation volume defined above.
The quark content used in the Wroblewski factor is determined at the moment
of {\it {chemical freeze-out}}, i.e.~from the hadrons and especially, hadronic
resonances, before they decay.
This ratio is thus not an easily measurable  observable
unless one can reconstruct all resonances from the final-state
particles.  The results are shown in Fig.~\ref{Wrob_composition} 
as a function of $\sqrt{s}$.

The solid line (marked ``sum'') in Fig.~\ref{Wrob_composition}
describes the statistical-model
calculations in complete equilibrium along the unified freeze-out
curve~\cite{CLEY98} and with the energy-dependent parameters $T$ and $\mu_B$.
From Fig.~\ref{Wrob_composition} we conclude that around 30 $A\cdot$GeV
laboratory energy the relative strangeness content in heavy ion
collisions reaches a
 clear and well pronounced maximum.
The Wroblewski factor  decreases towards higher incident energies
and reaches a limiting value of about 0.43.
For details see Ref.~\cite{max_strange}.

\begin{figure}
\begin{minipage}[t]{5.3cm}
\mbox{\epsfig{width=9.0cm,file=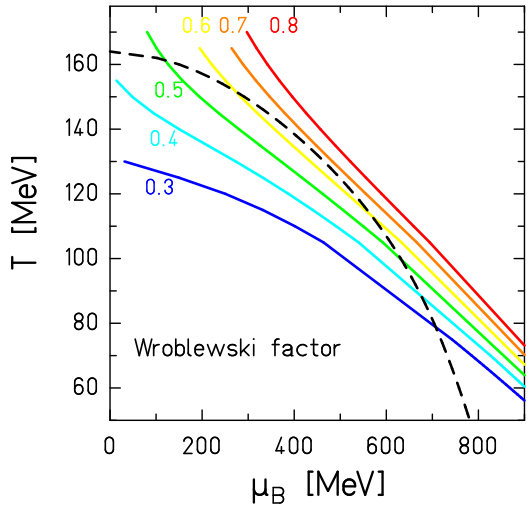}}
\end{minipage}
\hfill
\begin{minipage}[b]{7.8cm}
\vspace*{-4.0cm}
\caption{Lines of constant Wroblewski factor $\lambda_s$ (for
definition  see text) in the $T-\mu_B$ plane (solid lines)
together with the freeze-out curve (dashed line)~\protect\cite{CLEY98}.}
\label{Wrob_T_MUB}
\end{minipage}
\end{figure}

The appearance of the maximum can be traced  to the specific
dependence of $\mu_B$ and $T$ on the beam energy.
Figure~\ref{Wrob_T_MUB} shows
lines of constant $\lambda_s$  in the $T-\mu_B$
plane. As expected $\lambda_s$ rises with increasing $T$ for fixed
$\mu_B$.
Following the chemical freeze-out curve, shown as a dashed line in
Fig.~\ref{Wrob_T_MUB}, one can see that
 $\lambda_s$ rises quickly from SIS to AGS energies,
then reaches  a maximum at $\mu_B\approx 500$ MeV
and $T\approx 130$ MeV.
These freeze-out parameters correspond to
30 $A\cdot$GeV laboratory energy. At higher incident
energies the increase in $T$ becomes negligible but $\mu_B$ keeps
on decreasing and as a consequence $\lambda_s$ also decreases.

The importance of finite baryon density on the
behavior of $\lambda_s$ is demonstrated in  Fig.~\ref{Wrob_composition} showing
separately the  contributions to $\left<s\bar{s}\right>$
coming from strange baryons, from strange mesons and from hidden strangeness, 
i.e.~from hadrons  like $\phi$ and $\eta$.
As can be seen in Fig.~\ref{Wrob_composition},
the origin of the maximum in the Wroblewski ratio can be traced  to the 
contribution of strange baryons.
This channel dominates at low $\sqrt{s}$ and loses
importance at  high incident energies.
Even strange mesons exhibit a broad maximum. This is due to the
presence of associated production of e.g.~kaons together with
hyperons. 

 The energy dependence of the
$K^+/\pi^+$ ratio measured at midrapidity
is shown in Fig.~\ref{KP_PI_ratio}.
The model  gives an excellent description of the data, showing
a broad maximum at the same energy as the one
seen in the Wroblewski factor.
In general, statistical-model calculations
 should be compared with
4$\pi$-integrated results since strangeness does not have to be
conserved in a limited portion of phase space.
A drop in this
ratio for 4$\pi$ yields has been reported from preliminary results
of the NA49 collaboration  at 158 $A\cdot$GeV~\cite{blume}. This decrease
is, however, not reproduced by the statistical model
 without further modifications, e.g.~by introducing an additional
parameter $\gamma_s\sim 0.7$ \cite{becattini}.
This point might be clearer when data at other beam
energies  will become available.

\section{Summary}

Strange particle production in heavy ion collisions 
over a rather broad range of incident energies can be described by a statistical
model. The production of strange particles close to threshold 
requires a canonical formulation, i.e.~local strangeness conservation. 
This approach is able to explain
many features of $K^+$ and $K^-$ production at SIS energies.

While for $K^+$ production it remains open whether and how chemical
equlibrium can be reached, the situation is quite different for $K^-$. 
It is shown that the strangeness exchange process 
$\pi \Lambda \rightleftharpoons N + K^-$ is the dominant channel for 
$K^-$ production at SIS and likely also at AGS energies. This is demonstrated
by applying the corresponding law of mass action. Theoretical studies
confirm this interpretation. 

Using the energy dependence of the parameters $T$ and $\mu_B$
we have  shown
that the statistical-model description of relativistic heavy ion
collisions predicts that the yields of strange to non-strange
particles reaches  a well defined maximum near 30 GeV lab energy.
It is demonstrated that this maximum is due to the specific shape
of the freeze-out curve in the $T-\mu_B$ plane. In particular a
very steep decrease of the baryon chemical potential with
increasing energy causes a corresponding decline of relative
strangeness content in systems created in heavy ion
collisions above lab energies of 30 GeV. The saturation in $T$,
necessary for this result, might be connected to the fact that
hadronic temperatures cannot exceed the critical temperature
$T_c\simeq$ 170 MeV for the phase transition to the QGP as found
in solutions of QCD on the lattice.

In spite of the apparent success of the statistical models 
it should not appear the impression that these models describe everything. 
They describe yields, particle ratios. Looking at spectral shapes already 
the expansion dynamics shows up. The distribution of the particles in space 
is a very informative quantity as e.g.~\cite{Shin}. The description of this
quantity is beyond statistical models. 

It is a pleasure for me to thank for the stimulating collaboration
with  P. Braun-Munzinger, J.~Cleymans, K. Redlich, and the whole KaoS
Crew (I.~B\"ottcher, A.~F\"orster, E.~Grosse, P.~Koczo\'n, 
B.~Kohlmeyer, F.~Laue, M.~Menzel, L.~Naumann, F.~P\"uhlhofer, E.~Schwab, 
P.~Senger, Y.~Shin, H.~Str\"obele, F.~Uhlig, A.~Wagner, W.~Walu\'s).

\end{document}